\documentclass[pre,aps,twocolumn,showpacs,floatfix]{revtex4}

\usepackage[dvips]{graphicx}
\usepackage{amsmath}
\usepackage{amssymb}

\begin{document}
\bibliographystyle{prsty}

\title{L\'evy stable noise induced transitions: stochastic resonance, resonant activation and dynamic hysteresis}

\author{B. Dybiec}
\email{bartek@th.if.uj.edu.pl}
\affiliation{M. Smoluchowski Institute of Physics, and Mark Kac Center for
Complex Systems Research, Jagellonian University, ul. Reymonta 4, 30--059
Krak\'ow, Poland}

\author{E. Gudowska-Nowak}
\email{gudowska@th.if.uj.edu.pl}
\affiliation{M. Smoluchowski Institute of Physics, and Mark Kac Center for
Complex Systems Research, Jagellonian University, ul. Reymonta 4, 30--059
Krak\'ow, Poland}

\date{\today}

\begin{abstract}
A standard approach to analysis of noise-induced effects in stochastic dynamics assumes a Gaussian character of the noise term describing interaction of the analyzed system with its complex surroundings. An additional assumption about the existence of timescale separation between the dynamics of the measured observable and the typical timescale of the noise allows external fluctuations to be modeled as temporally uncorrelated and therefore white. However, in many natural phenomena the assumptions concerning the abovementioned properties of ``Gaussianity'' and ``whiteness'' of the noise can be violated. In this context, in contrast to the spatiotemporal coupling characterizing general forms of non-Markovian or semi-Markovian L\'evy walks, so called L\'evy flights correspond to the class of Markov processes which still can be interpreted as white, but distributed according to a more general, infinitely divisible, stable and non-Gaussian law. L\'evy noise-driven non-equilibrium systems are known to manifest interesting physical properties and have been addressed in various scenarios of physical transport exhibiting a superdiffusive behavior. Here we present a brief overview of our recent investigations aimed to understand features of stochastic dynamics under the influence of L\'evy white noise perturbations. We find that the archetypal phenomena of noise-induced ordering are robust and can be detected also in systems driven by non-Gaussian, heavy-tailed fluctuations with infinite variance. 
\end{abstract}

\maketitle

\section{Introduction\label{sec:introduction}}
Systems operating far from thermodynamic equilibrium are usually subjected to the action of noise which may profoundly influence their performance. Over the past two decades the extensive theoretical and experimental studies in physics, information theory and biology have documented various phenomena of noise-induced order, noise-facilitated kinetics or noise-improved signal transmission and detection. Almost all research in this field assumes that the noise process involved can be characterized by finite variance. Yet, in many situations the external non-thermal noise can be described by distribution of impulses following the heavy-tail stable law statistics of infinite variance. Notably, infinite variance does not exclude finite spread or dispersion of the distribution around its modal value. In fact,
for the family of $\alpha$-stable noises the interquantile distance can be used as a proper measure of the distribution width around the median.

Dynamical description of the system and its
surroundings is typically carried out within a stochastic picture
based on Langevin equations \cite{gammaitoni1998,anishchenko1999}. The basic equation of this type
reads
\begin{equation}
\dot{x}(t)=f(x,t) + \zeta(t),
\label{eq:generallangevin}
\end{equation}
where $f(x,t)$ is the deterministic (and possibly time-de\-pen\-dent) ``force'' acting on the system
and $\zeta(t)$ stands for the ``noise'' contribution describing
interaction between the system and its complex surrounding. If
 the noise can be considered as white and Gaussian, the above equation
gives rise to the classical Langevin approach used in
the analysis of Brownian motion. The whiteness of the noise
(lack of temporal correlations) corresponds to the existence
of time-scale separation between the dynamics of a relevant
variable of interest $x(t)$ and the typical time scale of the
noise. Hence white noise can be considered as a standard stochastic
process that describes in the simplest fashion the effects
of ``fast'' surroundings
\cite{shlesinger1995,nielsen2001}.

Although in various phenomena, the noise can be indeed interpreted as white (i.e. with stationary , independent increments), the assumption about its Gaussianity can be easily violated. The examples range from the description of the dynamics in plasmas \cite{chechkin2002b}, diffusion in energy space, self-diffusion in micelle systems, exciton and charge transport in polymers under conformational motion and incoherent atomic radiation trapping -- to the spectral analysis of paleoclimatic \cite{ditlevsen1999b} or economic data \cite{mantegna2000}, motion in optimal search
strategies among randomly distributed target sites \cite{Viswanathan1999}, fluorophore diffusion as studied in photo-bleaching experiments, interstellar scintillations \cite{boldyrev2003}, ratcheting devices \cite{dybiec2008d,delcastillonegrete2008} and many others \cite{kosko2001}.

The present work overviews
properties of L\'evy flights in external
potentials with a focus on astonishing aspects of noise-induced
phenomena \cite{jung1993,gammaitoni1995,schmitt2006} like resonant activation (RA), stochastic resonance (SR) and dynamic hysteresis \cite{dybiec2004,dybiec2004b,dybiec2006b,dybiec2007d,dybiec2007e}.
In particular, the persistency of the SR occurrence is examined within a continuous and a two-state description of the generic system \cite{dybiec2006b} composed of a test particle moving in the double well-potential and subject to the action of deterministic, periodic perturbations and $\alpha$-stable L\'evy type noises. In the same system the appearance of dynamic hysteresis is documented. Moreover, an archetypal escape problem over a fluctuating barrier is analyzed revealing the RA phenomenon in the presence of non-equilibrium L\'evy type bath.

\section{Model\label{sec:model}}

The model system is described by the following (overdamped) Langevin equation driven by a L\'evy stable, white noise $\zeta(t)$
\begin{equation}
\frac{dx(t)}{dt}=-V'(x,t)+\zeta(t).
\label{eq:langevin}
\end{equation}

\begin{figure}[!ht]
\begin{center}
\includegraphics[angle=0, width=7.0cm]{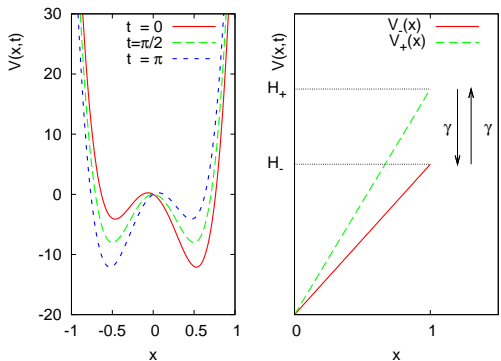}
\caption{The generic double-well potential $V(x,t)=-\frac{a}{2}x^2+\frac{b}{4}x^4+A_0x\cos\Omega t$ with $a=128,\;b=512,\;A_0=8$ for $t=\{0,\pi/2,\pi\}$ (the left panel) and the generic linear potential slope dichotomously switching between two configurations characterized by differend heights $H_\pm$ (right panel).}
\label{fig:model}
\end{center}
\end{figure}

To examine occurrence of stochastic resonance and dynamic hysteresis, the generic double-well potential with the periodic perturbation (see left panel of~Fig.~\ref{fig:model}) has been chosen
\begin{equation}
V(x,t) =-\frac{a}{2}x^2+\frac{b}{4}x^4 +A_0x\cos\Omega t,
\label{eq:potential}
\end{equation}
with $a=128,\;b=512,\;A_0=8$.

In turn, for inspection of resonant activation, the action of the potential $V(x,t)$ has been approximated by a linear slope switching dichotomously between two distinct configurations $V_\pm(x)=H_\pm x$ (see right panel of Fig.~\ref{fig:model}). The detection of RA has been carried for a relatively ``high'' barrier $H_+=8$, which under the normal diffusion condition guarantees the proper separation of time scales (the actual times of escape events and time of diffusive motion within the potential well \cite{hanggi1990,doering1992}) of the process. Following former studies on RA \cite{doering1992}, the lower barrier height has been chosen to take one of two values $H_-=-8$ or $H_-=0$. A particle has been assumed to start its motion at the reflecting boundary $x=0$ and continued as long as the position fulfilled $x(t)<1$.

It should be stressed that with generally non-Gaussian white noise the knowledge of the boundary location alone cannot specify in full the corresponding boundary conditions for reflection or absorption, respectively \cite{dybiec2006}. The trajectories driven by non-Gaussian white noise display irregular, discontinuous jumps. As a consequence, the location of the boundary itself is not hit by the majority of discontinuous trajectories. This implies that regimes beyond the location of the boundaries must be properly accounted for when setting up the boundary conditions. In particular, multiple recrossings of the boundary location from excursions beyond the specified state space have to be excluded. As it has been demonstrated elsewhere \cite{dybiec2006}, incorporation of L\'evy flights into kinetic description requires use of nonlocal boundary conditions which, for the case of absorbing boundary at, say, $x=1$ calls for extension of the absorbing regime to the semiline beyond that point, $x\geqslant 1$.

The barrier fluctuations causing alternating switching between the high ($H_+$) and low ($H_-$) barrier configurations have been approximated by the Markovian dichotomous noise \cite{horsthemke1984} with the exponential autocorrelation function $\langle [\eta(t)-\langle \eta(t)\rangle] [\eta(t')-\langle \eta(t')\rangle] \rangle =\frac{1}{4}(H_+-H_-)^2\exp(-2\gamma |t-t'|)$. Here $\gamma$ represents the transitions rates between $(\pm)$ and $(\mp)$ states, and both L\'evy and dichotomous noises have been assumed to be statistically independent.

The sample trajectories $x(t)$ have been obtained by a direct integration of the Langevin equation (\ref{eq:langevin}) using the standard techniques of integration of stochastic differential equations with respect to the L\'evy stable PDFs \cite{janicki1994,dybiec2004,dybiec2004b,dybiec2004c}
\begin{eqnarray}
x(t) & = &
-\int\limits_{t_0}^{t}V'(x(s),s)ds+\int\limits_{t_0}^{t}\zeta(s)ds \\
& \approx & -\int\limits_{t_0}^{t}V'((x(s),s))ds
+\sum\limits_{i=0}^{N-1}\Delta s^{1/\alpha}\zeta_i. \nonumber
\label{eq:lcalka}
\end{eqnarray}
Here $N\Delta s=t-t_0$ and
$\zeta_i$ is distributed according to the stable probability density function
$L_{\alpha,\beta}(\zeta;\sigma,\mu=0)$ whose corresponding Fourier transform for $\alpha\neq 1$ is
\begin{equation}
\phi(k) = \exp\left[ -\sigma^\alpha|k|^\alpha\left( 1-i\beta\mbox{sign}k\tan
\frac{\pi\alpha}{2} \right) \right],
\label{eq:charakt}
\end{equation}
while for $\alpha=1$ this expression reads
\begin{equation}
\phi(k) = \exp\left[ -\sigma|k|\left( 1+i\beta\frac{2}{\pi}\mbox{sign}k\log|k| \right) \right].
\label{eq:charakter}
\end{equation}

The parameter $\alpha\in(0,2]$ denotes the stability index, yielding the asymptotic long tail power law for the $\zeta$-distribution, which for $\alpha<2$ is of the $|\zeta|^{-1-\alpha}$ type. The parameter $\sigma$ ($\sigma\in(0,\infty)$) characterizes the scale whereas $\beta$ ($\beta\in[-1,1]$) defines an asymmetry (skewness) of the distribution
\cite{janicki1994,nolan2002}.

\begin{figure}[!ht]
\begin{center}
\includegraphics[width=8.0cm]{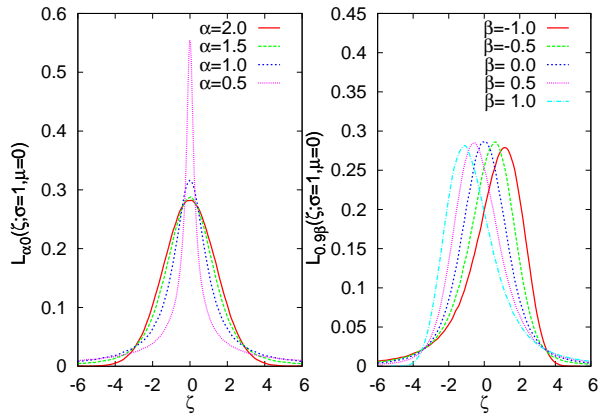}
\caption{Symmetric stable distributions for $\alpha=2. 0, 1. 5, 1. 0, 0. 5$ with $\beta=0$ (from the bottom to the top) (left panel) and asymmetric stable densities for $\beta=1.0, 0.5, 0.0, -0.5, -1.0$ with $\alpha=1.5$ (from left to right) (right panel).}
\label{fig:densities}
\end{center}
\end{figure}

Exemplary probability density functions for symmetric ($\beta=0$) and skewed ($\beta\neq 0$) cases have been displayed in Fig.~\ref{fig:densities}.

\section{Results \label{sec:solution}}

\subsection{Stochastic resonance and dynamic hysteresis \label{sec:sr}}

The response of the ensemble average coordinate $\langle x(t) \rangle$ to external periodic perturbations (cf. Eq.~(\ref{eq:langevin})) gives rise to the phenomenon of ``dynamic hysteresis''. In order to register the behavior, we have performed an analysis of trajectories $x(t)$ based on a two-state approximation. For this purpose, we have defined an occupation probability of being in the left (with the respect to the barrier position $x=x_b$) or right state according to the definition of the occupation probabilities:
\begin{equation}
p_{\mathrm{left}}(t)=\mathrm{Prob}[x(t)<x_b]=\int\limits^{x_b}_{-\infty}p(x,t)dx=1-p_{\mathrm{right}}(t).
\end{equation}

\begin{figure}[!ht]
\begin{center}
\includegraphics[angle=0, width=7.0cm]{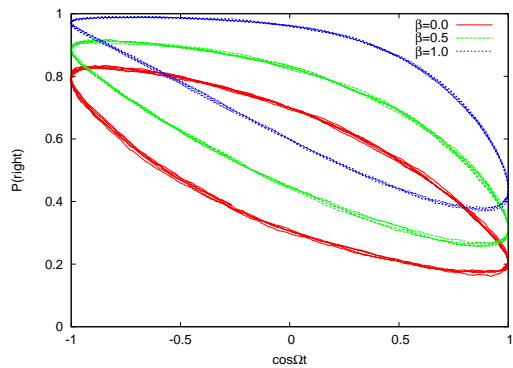}
\caption{Dynamic hysteresis loops for various $
\beta$ with fixed $\alpha=1.9$. The time step of the integration $\Delta t=10^{-4}$. Results were averaged over $N=10^3$ realizations. Initial conditions $x(0)$ were sampled from the interval $[-1.25,1.25]$. The particle is moving in the modulated double-well potential (\ref{eq:potential}) with $a=128,\;b=512,\;A_0=8,\;\Omega=1,\;\sigma=\sqrt{2}$.}
\label{fig:sh_a19}
\end{center}
\end{figure}

Note that the relative occupation of either one of those states can be altered not only by the modulation of the potential, but also by an appropriate tuning of the additive noise parameters. When plotted as a function of the periodic driving $\cos\Omega t$, the occupation probability
$p_{\mathrm{right}}(t)$ exhibits a characteristic hysteresis loop. Quite obviously, the nonzero skewness parameter $\beta$ introduces asymmetry to the dynamic hysteresis loops, cf. Fig.~\ref{fig:sh_a19}. As can be inferred from Fig.~\ref{fig:sh_a19}, the positive $\beta$ ``biases'' the motion, causing the trajectories to stay more likely to the right of the barrier top $x=0$.

The overall behavior of the dynamic hysteresis loops
\begin{equation}
HL\equiv \int\limits_{t=0}^{2\pi/\Omega}  \langle x(t) \rangle d(\cos\Omega t)
\end{equation}
under the $\alpha$-stable noises could be deduced from the inspection of the trajectories of the process presented in Fig.~\ref{fig:strajectory}. For the symmetric stable noise, i.e. $\beta=0$, the process spends, on average, the same amount of time in the left/right states and consequently, occupations of both states are equal. For non-zero and increasing $|\beta|$ a larger asymmetry in the distribution of residence times in the right/left states is registered. The asymmetry of the occupation probability in either one of the states is reflected in the shape of the dynamic hysteresis loop, which with an increasing $|\beta|$ becomes distorted into the direction determined by the sign of the skewness parameter.

\begin{figure}[!ht]
\begin{center}
\includegraphics[angle=0, width=8.0cm]{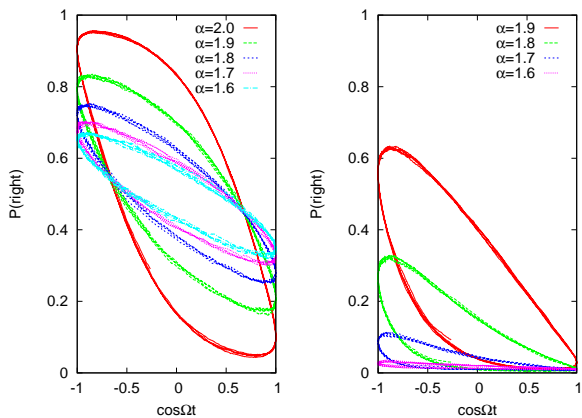}
\caption{Dynamic hysteresis loops for various $\alpha$ with $\beta=0$ (the left panel) and $\beta=-1$ (the right panel). Parameters of the simulation like in Fig.~\ref{fig:sh_a19}.}
\label{fig:sh_alpha}
\end{center}
\end{figure}

\begin{figure}[!ht]
\begin{center}
\includegraphics[angle=0, width=7.0cm]{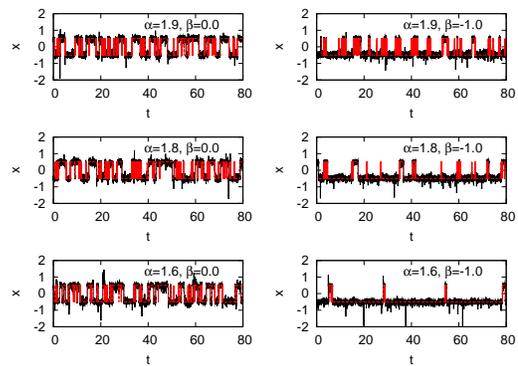}
\caption{Trajectories of the $\alpha$-stable process for the two-state and the continuous model constructed by use of Eq.~(\ref{eq:langevin}). The time step of the integration $\Delta t=10^{-5}$, the scale parameter $\sigma=\sqrt{2}$. The frequency of the cycling voltage $\Omega=1$. Noise parameters as indicated in figures.}
\label{fig:strajectory}
\end{center}
\end{figure}

The probability of finding the process in the right/left state depends strongly on the stability index $\alpha$, cf. Fig.~\ref{fig:sh_alpha}.
For symmetric noises ($\beta=0$) the area of the hysteresis loop decreases with decreasing stability index $\alpha$, see Fig.~\ref{fig:sh_alpha}. This observation is a direct consequence of a heavy-tailed nature of the noise term in Eq.~(\ref{eq:generallangevin}). With decreasing $\alpha$, larger excursions of the particle are possible and these occasional jumps of trajectories may be of the order of, or even larger than the distance separating the two minima of the potential $V(x,t)$. A combined interplay of both noise parameters $\alpha,\beta$ may result in a permanent locking of the process in one of its states, see right panel of Fig.~\ref{fig:sh_alpha}.

\begin{figure}[!ht]
\begin{center}
\includegraphics[angle=0, width=8.0cm]{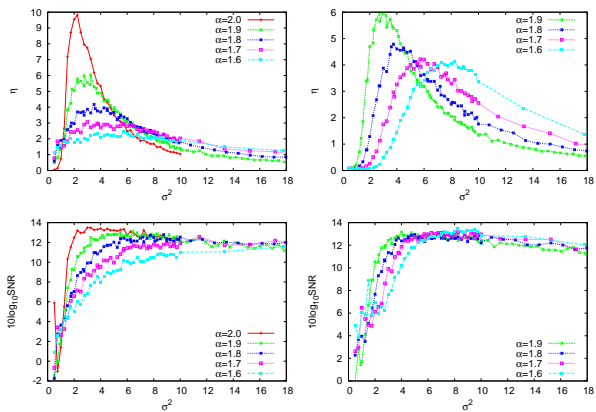}
\caption{SPA in arbitrary units (the top panel) and SNR (the bottom panel) for the perturbed generic double-well potential driven by additive L\'evy noises with $\beta=0$ (the left panel), $\beta=-1.0$ (the right panel) and various $\alpha$. The simulation parameters like in Fig.~\ref{fig:strajectory}. Lines are drawn to guide the eye.}
\label{fig:genericpower}
\end{center}
\end{figure}

In order to quantify the SR phenomenon \cite{gammaitoni1998,talkner2005} we have used the standard measures of the signal-to-noise ratio SNR
and the spectral power amplification $\eta$ \cite{gammaitoni1998}. The SNR is
defined as a ratio of a spectral content of the signal in
the forced system to the spectral content of the noise $\zeta(t)$:
\begin{equation}
SNR=2\left[\lim_{\Delta \omega\rightarrow 0}\int\limits^{\Omega+\Delta
\omega}_{\Omega-\Delta \omega} S(\omega)d\omega\right]/S_N(\Omega).
\end{equation}
Here $ \int^{\Omega+\Delta \omega}_{\Omega-\Delta \omega} S(\omega)d\omega$
represents the power carried by the signal, while $S_N(\Omega)$
estimates the background noise level. In turn, the spectral power amplification ($\eta$)
is given by the ratio of the power of the driven oscillations to that of the driving signal
at the driving frequency $\Omega$.
Closer examination of these quantifiers reveals that $\eta$ and SNR behave in a typical way both in the continuous model, cf. Fig.~\ref{fig:genericpower},
and in its two-state analogue (results not shown). The spectral amplification $\eta$ has a characteristic bell-shape form indicating
detection of stochastic resonance within the interval of suitably chosen noise intensity $\sigma^2$.
For example, if the stability index is set to $\alpha=1.9$ and trajectories are simulated with extremely weak noise intensities $\sigma^2$, the periodic signal is not well separated from the noisy background and consequently SNR stays negative.

However, SNR becomes positive with increasing noise intensity what indicates emerging separation of the signal from the noisy background. In Fig.~\ref{fig:genericpower} results for $\eta$ and SNR analysis (in a continuous SR model) with various stability index $\alpha$ are presented. Note, that for a given $\alpha$ the power spectra for $\pm\beta$ are the same. Therefore, stochastic resonance quantifiers derived from the power spectra are equivalent for $\pm\beta$ with the same value of $\alpha$.

Switching from the continuous model to the two-state approximation (cf. Fig.5) does not change the qualitative behavior of the SR
quantifiers.\footnote{In fact, some
changes are observed but in the parameter region where barrier crossing events are not recorded, i.e. at relatively small noise intensity $\sigma^2$ with large $\alpha$ ($\alpha\approx 2$)} However, the slope of the decaying part of the signal-to-noise ratio is more flat for the two-state description than for the continuous model (results not shown).

Our numerical analysis implies that $\eta$ is more sensitive to the variation of noise parameters than SNR, see Fig.~\ref{fig:genericpower}. At decreasing values of the stability index $\alpha$, the maximum of $\eta(\sigma^2)$ drops and shifts towards higher values of the noise intensity. Obviously, the decrease of $\alpha$ weakens the stochastic resonance and reduces system performance. The diminishment of spectral amplification for $\alpha<2$ indicates that the input signal is worse reproduced in the recorded output. This behavior is easier detectable for symmetric noises (cf. left panel of Fig.~\ref{fig:genericpower}) than for asymmetric ones (see the right panel) and
 can be readily explained by the analysis of exemplary trajectories, see Fig.~\ref{fig:strajectory}. Sharp spikes clearly visible in the right corner panel of Fig.~\ref{fig:strajectory} are due to the heavy-tailed nature of the L\'evy stable distribution and they are more pronounced the smaller the index $\alpha$ becomes. Their presence indicates
that for a sufficiently small $\alpha$ the trajectory becomes discontinuous and, on average, switches between left and right wells of the potential $V(x,t)$ are realized by sudden long-jump escape events fairly independent of the periodic driving.
Although the SNR is less sensitive to variations in noise parameters (cf. Fig.~\ref{fig:genericpower}), the shape of this function flattens for decreasing values of $\alpha$ thus hampering detection of the resonant value of the noise optimal intensity for which the SR phenomenon is most likely perceived.

\subsection{Resonant Activation\label{sec:ra}}
As already discussed in the preceding sections, the stochastic kinetics driven by additive non-Gaussian stable white noises is very different from the Gaussian case. For $\alpha<2$, a test particle moving in the linear potential can change its position via extremely long, jump like excursions. This in turn requires the use of nonlocal boundary conditions in evaluation of the mean first passage time (MFPT) \cite{dybiec2006,dybiec2007,zoia2007,koren2007}.
In this paragraph the above issue is taken care of when generating first passage times (FPTs) by Monte Carlo simulations. Simulated trajectories are representative for a motion of a test particle over the interval $[0,1]$ (see right panel of Fig.~\ref{fig:model}) influenced by independent dichotomous switching of the potential slope and subjected to additive white L\'evy noise. A particle starts its motion at $x=0$ where the reflecting boundary is located. The absorbing boundary is located at $x=1$ meaning, that the whole semi-axis $[1,\infty)$ is assumed absorbing, yielding zero PDF $p(x,t)=0$ for all $x\geqslant 1$.

\begin{figure}[!ht]
\begin{center}
\includegraphics[angle=0, width=8.0cm]{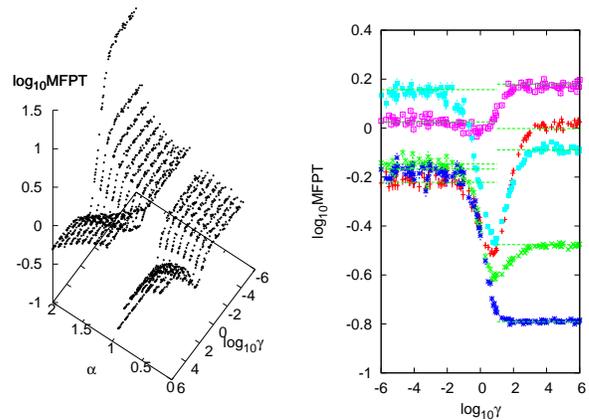}
\caption{$\mathrm{MFPT}(\alpha,\gamma)$ for $H_\pm=\pm8$ with $\beta=1,\sigma=1$ (the left panel) and sample cross-sections $\mathrm{MFPT}(\gamma)$ for various $\alpha$: `$+$': $\alpha=0.2$; `$\times$': $\alpha=0.8$; `$\ast$': $\alpha=0.9$; `$\square$': $\alpha=1.1$ and `$\blacksquare$': $\alpha=1.7$ (the right panel). The results were calculated by direct integration of Eq.~(\ref{eq:langevin}) with the time step $\Delta t=10^{-5}$ and averaged over $N=10^3$ realizations. Horizontal lines represent asymptotic values of $\mathrm{MFPT}(\gamma\to 0)$ and $\mathrm{MFPT}(\gamma\to\infty)$. They have been evaluated by use of the Monte Carlo method with $ \Delta t=10^{-5}$ and averaged over $N=5\times10^3$ realizations. Error bars represent standard deviation of the mean.}
\label{fig:skewedm88}
\end{center}
\end{figure}

From the ensembles of collected first passage times we have evaluated the mean values of the distributions and our findings for MFPT are presented in Fig.~\ref{fig:skewedm88}. Integration of Eq.~(\ref{eq:langevin}) was performed for a series of descending time steps of integration $\Delta t=\{10^{-2},10^{-3},10^{-4},10^{-5}\}$ to ensure that the results are self consistent.

The left panel of Fig. \ref{fig:skewedm88} displays behavior of derived MFPTs as a function of the stability index $\alpha$ and the rate of the barrier modulation $\gamma$. In the right panel sample cross-sections of the surfaces MFPT($\alpha,\gamma$) are drawn. For $\alpha=1$, or adequately $\alpha\approx 1$, the procedure of simulating skewed ($\beta\neq 0$) stable random variables becomes unstable.  This can be well explained by examining the form of the additive noise-term characteristic function $\phi(t)$ - see Eqs.~(\ref{eq:charakt}) and (\ref{eq:charakter}). The exponential functions are no longer continuous functions of the parameters and exhibit discontinuities when $\alpha=1, \beta\neq0$.
Therefore, for the clarity of presentation, these parameter sets have been excluded from consideration.

The depicted results indicate appearance of the RA phenomenon which is best visible when the potential barrier is switching between two barrier configurations characterized by $H_\pm=\pm8$.
However, the dependence of RA on noise parameters is highly nontrivial. The overall tendency in kinetics is similar to cases described in former paragraphs: heavier tails ($\alpha<2$) in distribution of noise increments $\zeta(t)$ result in stronger discontinuities of trajectories causing the RA effect to become inaudible and fading gradually. Notably, for totally skewed additive noise $\beta=1$ which acts in favor of the motion to the right, the RA seems to disappear for $\alpha=0.9$. It reappears again for smaller values of $\alpha$, when the driving L\'evy white noise becomes a one sided L\'evy process with strictly positive increments which tend to push trajectories to the right.

\begin{figure}[!ht]
\begin{center}
\includegraphics[angle=0, width=8.0cm]{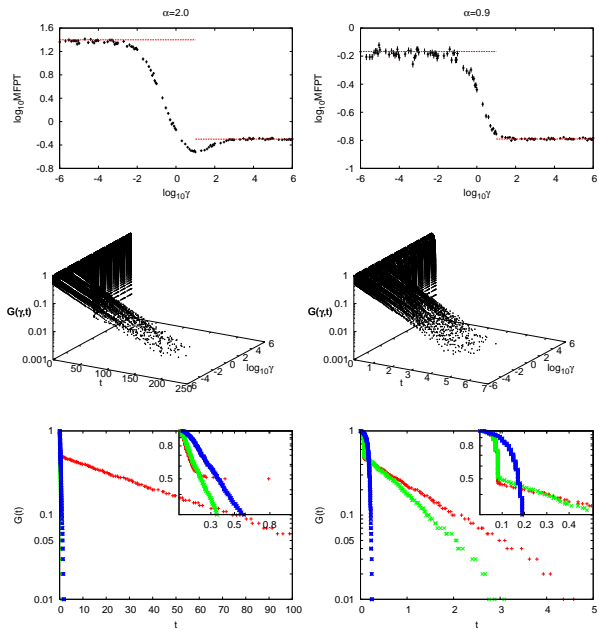}
\caption{Comparison of the behavior of MFPTs for $\alpha=2$ (left panel) and $\alpha=0.9$ (right panel) for $H_\pm=\pm8$. MFPTs' curves (upper panel), survival probability surfaces $G(\gamma,t)$ (middle panel) and sample cross-sections of $G(\gamma,t)$ surface (lower panel) for small ('$+$'), resonant ('$\times$') and large ('$*$') $\gamma$-values are depicted. Simulations parameters as in Fig.~\ref{fig:skewedm88}. Note the log scale on z-axis (middle panel) and y-axis (lower panel).}
\label{fig:survm88}
\end{center}
\end{figure}

To further examine the character and distribution of escape events from $x=0$, we have analyzed survival probabilities $G(\gamma, t)$ constructed from generated trajectories at fixed values of frequencies $\gamma$. For better comparison with the Gaussian RA scenario, the summary of results is displayed in Fig.~\ref{fig:survm88} with sets of data relating to $\alpha=2$ and $\alpha=0.9$.
The left and right panels of Fig.~\ref{fig:survm88} present behavior of MFPT as the function of the barrier modulation parameter $\gamma$, the survival probability $G(\gamma,t)$ and exemplary cross-sections of the survival probability surface. The insets depict behavior of the survival probability $G(\gamma,t)$ at short time scales (small $t$).

At low rates of the barrier switching process (small $\gamma$), two distinct time scales of the barrier crossing events can be observed. The fast time scale corresponds to passages over the barrier in its lower state, while the large time scale is pertinent to the slower process, i.e. the passages over the barrier in its higher energetic state. This effect is well pronounced after the RA phenomenon sets up. The presence of the two time scales for  small $\gamma$ and only one time scale for large $\gamma$ explains the asymptotic behavior of MFPTs. Namely, for low values of the switching rate $\gamma$, MFPT is an average value of MFPTS over both configuration of the barrier. With increasing frequency $\gamma$ the distinguishable time scales coalesce and disappear. This is due to the fact that the barrier changes its height multiple times during the particle's motion so that resulting MFPT describes kinetics over the average potential barrier.

\section{Discussion\label{sec:discussion}}

We have examined the influence of various types of stochastic $\alpha$-stable drivings on dynamic properties of a generic two-state model system. Although our primary interest was to understand the effects of L\'evy-type drivings with the stability index $0<\alpha<2$, the numerical analysis as applied in these studies remains valid for any set of parameters characterizing $\alpha$-stable excitations. 

The response to external or/and parametric perturbations has been examined by analyzing qualitative changes in system's dynamic behavior as expressed in the onset of resonant activation, stochastic resonance and dynamic hysteresis. Due to the inherent symmetry of the potential, the SR quantifiers constructed at a given value of the stability index $\alpha$ have been the same for $\pm\beta$. On the other hand, the asymmetry of the driving noise has been shown to influence strongly the population of states and therefore has affected mostly the appearance and performance of the dynamic hysteresis.

The system efficiency has been described by standard SR measures, i.e. signal-to-noise ratio (SNR) and spectral power amplification ($\eta$). Those quantifiers have been shown to behave in a typical way. In other words, by tuning the noise intensity $\sigma^2$, the maximum in the  SNR($\sigma^2$) and $\eta(\sigma^2)$ could be detected, thus proving that stochastic resonance is a robust phenomenon which may be observed also in systems subjected to the action of impulsive jump-noise stochastic processes. Decrease in stability index $\alpha$ results in larger jump-like excursions of the process $x(t)$ and consequently causes weakening of SR. Out of two SR measures, the spectral power amplification has turned out to be  more sensitive to variations of $\alpha$  than SNR. A more pronounced drop in system efficiency has been observed for symmetric noises with $\beta=0$. 

The nonequilibrated, non-thermal L\'evy white noise affects also a paradigm scenario  of escape kinetics. By numerically implementing the set up boundary conditions for the problem, we have investigated the statistics of escape times over the dichotomously fluctuating barrier. The manifestation of the RA phenomenon  has been analyzed within a certain frequency $\gamma$ regime pointing that by a continuous readjustment of the external noise paramaters the resonant activation can be either suppressed or re-induced.

L\'evy white noise with $\alpha\neq 2$ extends a standard Brownian noise to a vast family of impulsive jump-like stochastic processes. Our studies  document that dynamical systems driven by such sources can also benefit and display a noise-enhanced order.

\begin{acknowledgements}
The research has been supported by the Marie Curie TOK COCOS grant (6th EU
Framework Program under Contract No. MTKD-CT-2004-517186) and European Science
Foundation (ESF) via `Stochastic Dynamics: fundamentals and applications'
(STOCHDYN) program. \ Additionally, BD acknowledges the support from the
Foundation for Polish Science.
\end{acknowledgements}


\end{document}